\documentclass[onecolumn]{aa}
\usepackage{graphicx}

\begin{document}

\title {Tidal dwarf galaxies candidates in a sample of minor merger of galaxies.}

\author{D. L. Ferreiro\inst{1} \fnmsep\thanks{Visiting Astronomer at the Cerro Tololo Iter-American
Observatory (CTIO).}, 
	M. G. Pastoriza\inst{2} 
	\and
	M. C. Rickes\inst{2}  }

\institute{IATE, Observatorio Astron\'omico, Universidad Nacional de
    C\'ordoba, Laprida 854, 5000, C\'ordoba, Argentina\\
	\email{diegof@mail.oac.uncor.edu}
	\and
	Departamento de Astronomia, IF-UFRGS, C.P. 15051, CEP
    91501-970, Porto Alegre, RS, Brazil\\
	\email{miriani.pastoriza@ufrgs.br}\\
	\email{maurogr@if.ufrgs.br}  }

\date{Received XXXX; accepted XXXX}

\abstract{We present in this paper a list of candidates tidal dwarf galaxies selected among a sample of 117 HII region observed in 11 minor mergers. The classification of the HII regions was performed using the blue absolute magnitude (M$_B$ $<$ -15), H$\alpha$ luminosity ($\cal L$$(H\alpha)$ $>$ 10$^{39}$ erg s$^{-1}$) and the star formation rate ($SFR_{H\alpha}$ $>$ 0.4 M$_{\odot}$ y$^{-1}$) as parameters. The total number of UV photons, the number of the star of each spectral type and the total stellar mass of the cluster were computed for different models adopting Salpeter's IMF with $\alpha = -2.65$ for the ionizing cluster. The ionizing cluster model which better reproduce the observed properties are supermassive clusters with mass greater than 10$^6 \leq M/M_{\odot} \leq 10^7$ .   
 }

\authorrunning {D. Ferreiro, M. Pastoriza \& M. Rickes}

\titlerunning {Tidal dwarf galaxies candidates}

\maketitle

\section{Introduction}
Zwicky (1956) and Schweizer (1978) were the first to propose that collisions between giant galaxies may eject dwarf galaxies to intergalactic/intercluster space. The stars and gas may be gravitationally pulled out of their parent galaxies during tidal encounters, forming rings, tails and bridges. The amount of matter lost during the outflow can be as large as 1/3 of the mass in the pre-encounter disk (Duc \& Mirabel 1999). \\

Models for the formation of tidal Dwarf Galaxies (hereafter TDGs) put forward two mechanisms (Fritze et. al 1998):
\begin{itemize} 
\item\ Stellar-dynamical models reveal local concentrations of stars along stellar tidal tails torn out from the disk of an interacting spiral (Barnes \& Hernquist 1992). Gas, if present, may then fall into the potential well defined by the disk.
\item\ Hydro-dynamical models show local instabilities of gas along gaseous tidal tails that give rise to Super-Giant Molecular Clouds, which then may initiate a Burst of Star Formation (Elmegreen et al. 1993). Some stars, if present, might then fall into the potential defined by the gaseous component.
\end{itemize}

Blue knots can be seen along the tidal tails. These condensations of gas and stars may detach from the system in dwarf galaxies or star clusters. But what are TDGs? Weilbacher and Duc (2001) define a TDG as a self-gravitating entity of dwarf-galaxy mass built from tidal material expelled during interactions. These galaxies are a new class of ``recycled" objects with some properties similar to classical dwarf irregulars (dIrrs) and blue compact dwarf galaxies (BCDGs). Their spectra present optical emission lines, typical of HII regions ionized by OB stars younger than 10 Myrs. The TDGs have blue colors as a result of active starburst. Two types of stars can be found in TDGs: old and young ones. The first ones were made in the parent galaxies and pulled out from the disk by the tidal forces, and the second ones were formed recently by collapse from the expelled HI clouds. Star formation in these regions occur at rates which might reach 0.1 M$_{\odot}$ yr$^{-1}$ (Duc \& Brinks 2001). Since TDGs contain huge HI reservoirs, typically 10$^9$M$_{\odot}$, one should expected that the relative importance of an older stellar population would decrease with time as star formation proceed.\\

Substantial quantities of molecular gas have been detected in the form of CO emission (Braine et al. 2000), assuming that the conversion factor between the molecular gas and CO emission is similar to the Galactic value, a large amount of molecular gas should be present in TDGs. This fact together with the spatial and dynamical coincidence of CO emission along with H$\alpha$ emission suggest that we are in the presence of the molecular gas, being transformed into atomic gas and subsequently in stars. (Lisenfeld et al. 2001).\\

The most probable location of the survivor TDGs is towards the tip of the tidal tails, since those formed close to the progenitor would quickly disappear destroyed by the gravitational potential of the parent galaxy (Bournaud et al. 2003). One should however consider that in the case of interactions between unequal mass galaxies, the length of the tidal tail should be much smaller. Furthermore, dwarf candidates would be located much closer to the nuclei of the interacting pair.\\

In the previous paper we have estimated the basic photometric parameters of the HII regions of eleven minor mergers of galaxies, from H$\alpha$+N[II] images (Ferreiro \& Pastoriza 2004, hereafter FP2004). Most of the detected HII regions of the sample were formed between 3.6 to 13.7 Myrs. ago with an average of (6.3$\pm$0.7) Myrs. We have found that the HII region properties, luminosity, sizes and ages are similar in both components. The HII regions have log(H$\alpha$+[NII]) luminosity between 38.6 and 41.7 and the HII region luminosity function for the whole sample fits a power law of index  $\alpha$ = --1.33 showing an excess of very luminous objects compared with normal galaxies, which have power law index of 2. \\

The goal of this paper is to study the nature of the HII regions of the sample of minor mergers searching for the presence of TDGs. The paper is organized as follows: In Sect.~\ref{sam obs red} we describe the sample, summarize the observations and briefly describe the data reductions. Sect.~\ref{candidates TDG} presents the classification of the possible nature of HII regions. Sect.~\ref{modeling the Ionizing cluster} we model the ionizing cluster. In Sect.~\ref{discussions and conclusions} we present the final remarks. \\

\section{Sample, observation and data reduction}
\label{sam obs red}

The HII regions studied in this paper were detected in H$\alpha$ images of eleven minor mergers of galaxies\footnote{The condition for the minor merger is $\cal M$$_{(secondary)}$/$\cal M$$_{(primary)}$ $< 0.2$}. These pairs of galaxies were selected from the Catalogue of Arp \& Madore (1987) and previously studied by FP2004.\\

Narrow band H$\alpha$ images and continuous (6400 \AA) were observed in July 1999 using Tektronix 2048 $\times$ 2048 CCD attached to the 0.90 m Cerro Tololo Inter-American Observatory Telescope. The pixel scale is 0.396 arcsec. Seeing conditions were good (1".1 -- 1".4). \\

All images were reduced following the standard procedures using the IRAF package. These procedures are bias subtraction, flat field normalization, sky subtraction, cosmic ray removal, seeing estimation, extinction correction and calibration with standard stars. Estimates of the accuracy in the calibrations are $\pm$ 0.04 mag in B, $\pm$ 0.06 mag in (B-V) and $\pm$ 0.06 mag in (V-I). Throughout this paper, a Hubble constant of $H_0$ = 75 $km$ $s^{-1} Mpc^{-1}$ is adopted.\\

For each H$\alpha$ image a suitable H$\alpha$ continuum image was subtracted. Prior to this subtraction a careful alignment of the images was made. The typical accuracy was better than 0.5 pixel. When the images had different seeing, the ones with better seeing were convolved with a Gaussian function in order to match the image with the worst seeing.\\

The H$\alpha$+[NII] images\footnote{Image B, V, I and H$\alpha$ of the sample are available at http://www.edpsciences.org/aa.} were used to identify HII regions for each galaxy and measured the CCD positions (X and Y) (FP2004).\\

The size of the HII region was defined as the equivalent radius  $r_{eq}$ = ($\cal A$/$\pi$)$^{0.5}$ where $\cal A$ is the area inside the isophotal level that has an intensity value of 10\% of the central intensity of the HII region.  \\

The H$\alpha$+[NII] flux and luminosity were determined integrating the intensity inside a diaphragm of the equivalent radius. The flux calibration was performed through calibrated spectra of some of the observed HII regions. For each H$\alpha$ image a suitable H$\alpha$ continuum image was also observed. Therefore we were able to calculate the equivalent width $EW(H\alpha+[NII])$ for the individual HII regions.\\

We have also measured for the HII regions: B magnitude, B-V and V-I colours inside the equivalent radius. We have estimated the internal reddening for each galaxy from the line intensity ratio H$\alpha$/H$\beta$ observed in the nuclear region spectra given by Pastoriza, Donzelli \& Bonatto (1999). The internal reddening derived from the H$\alpha$/H$\beta$ ratio of the nuclear spectrum are in the range of $0.25\ < E(B-V)\ <\ 0.96$. We consider this reddening as an upper limit for the knots.  \\

In table~\ref{TDG-cand} we present for the star-forming regions listed in order of decreasing brightness the following parameters: identification of the regions, the Arp galaxy to which they belong, the blue absolute magnitude $M_B$, the integrated $(B-V)$ and $(V-I)$ colours, the $H\alpha$ luminosity ($\cal L\rm (H\alpha)$), the number of ionizing photons $Q(H)$, the age of the regions and the SFR($H\alpha$). 
The number of ionizing photons $Q(H)$ for the observed regions (column 7), were estimated from the $H\alpha$ luminosity $\cal L\rm (H\alpha)$ (column 6) corrected by the contribution of the [NII] $\lambda\lambda$6548, 6584 emission. This contribution was assumed to be 20\% in all the regions. This correction was taken from spectroscopic observations of galactic HII regions (Girardi et al. 1997). The number of ionizing photons was estimated by the Eq.~\ref{eq1} of Osterbrock (1989):

\begin{eqnarray}
Q(H) = \frac{L_{H\alpha}}{h\nu_{\alpha}} \frac{\alpha_B(H^0,T)}{\alpha_{H\alpha}(H^0,T)}10^{c(H\alpha)}
\label{eq1}
\end{eqnarray}

where $\alpha_B(H^0,T)$ is the coefficient of recombination added on all levels of energy except the fundamental one $H\alpha(H^0,T)$ is the coefficient of recombination in $\alpha$ and the term $10^{c(H\alpha)}$.  
The age of the regions, listed in column 8 was estimated from the equivalent width $EW(H\alpha+[NII])$ and the $B-V$ and $B-I$ colours using the synthesis models of Leitherer et al. (1999) (See FP2004).  
We have calculated the SFR($H\alpha$) in the observed HII regions using the following equation~\ref{SFR(Ha)} (Kennicutt 1998): 

\begin{equation} 
SFR_{H\alpha} = 7.9\ 10^{-42}\ \cal L\rm(H\alpha)\ M_\odot\ yr^ {- 1} 
\label{SFR(Ha)}
\end{equation} 

This equation was derived using the calibrations of Kennicutt et al (1994) and Madau et al (1998), assuming stars with solar abundances and Salpeter's Initial Mass Function (IMF) with a mass interval of $0.1\ M_\odot <M_\star\ < 100\ M_\odot$. \\

\section{Nature of the star formation regions}
\label{candidates TDG}

As it was described in FP2004, a total of 117 HII regions were identified in the 11 pairs of minor merger. In this section we classify these regions using three parameters:\\

\begin{itemize} 
\item\ The blue absolute magnitude $M_B$ 
\item\ The $H\alpha$ luminosity 
\item\ The star formation rate $SFR_{H\alpha}$. 
\end{itemize}

The blue absolute magnitude M$_B$ listed in Table~\ref{TDG-cand} is the integrated blue magnitude of the whole region inside of the equivalent area corrected by the contribution of the underlying disk (see FP2004). The correction was estimate from the luminosity profile decomposition in bulge and disk. This contribution is smaller than 1\% for most of the regions therefore negligible inside the errors ($\sim 10\%$).\\

Using the blue luminosity the observed HII regions were classified in three categories (Duc et. al 2004):
\begin{itemize} 
\item\ {$Super\ Stellar\ Cluster\ (SSC)$:\\ 
These regions are weaker than  $M_B> -12\ mag $. This brightness is characteristic of HII regions observed in the arms of normal Sa/Sb type galaxies (Bresolin \& Kennicutt 1997).}\\

\item\ {$Giant\ HII\ Complex\ (GHIIC)$: \\
Objects with brightness between $-15\ mag <M_B < -12\ mag$ which corresponds to the interval of the brightest HII region observed in normal Sc type galaxies (Bresolin \& Kennicutt 1997).}\\

\item\ {$Tidal\ Dwarf\ Galaxies\ Candidates\ (TDGs)$:\\
Regions with luminosity brighter than $M_B < -15\ mag$.}\\
\end{itemize}

\begin{figure*} 
\begin{center} 
\includegraphics[width=8cm,height=8cm]{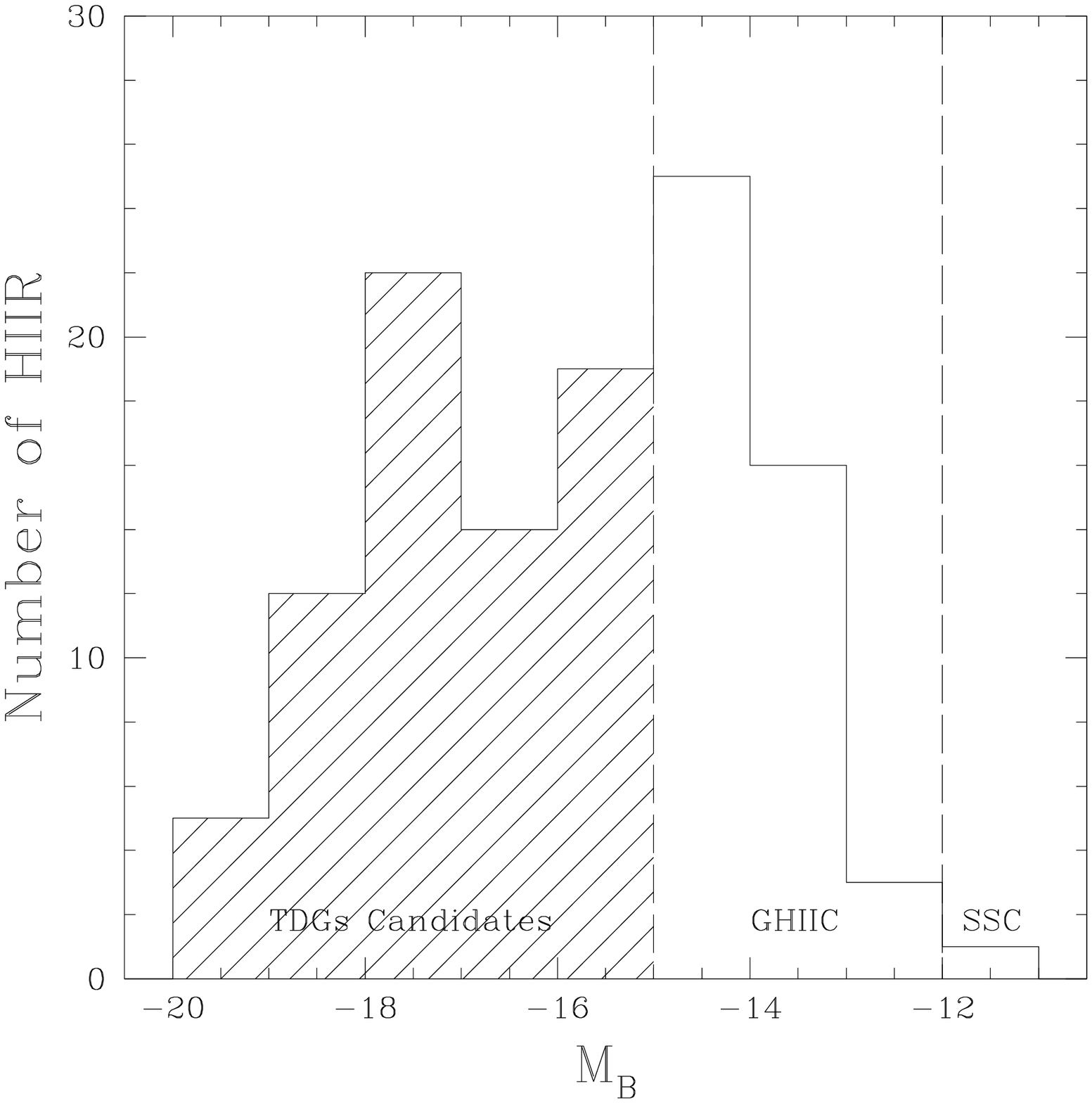}
\includegraphics[width=8cm,height=8cm]{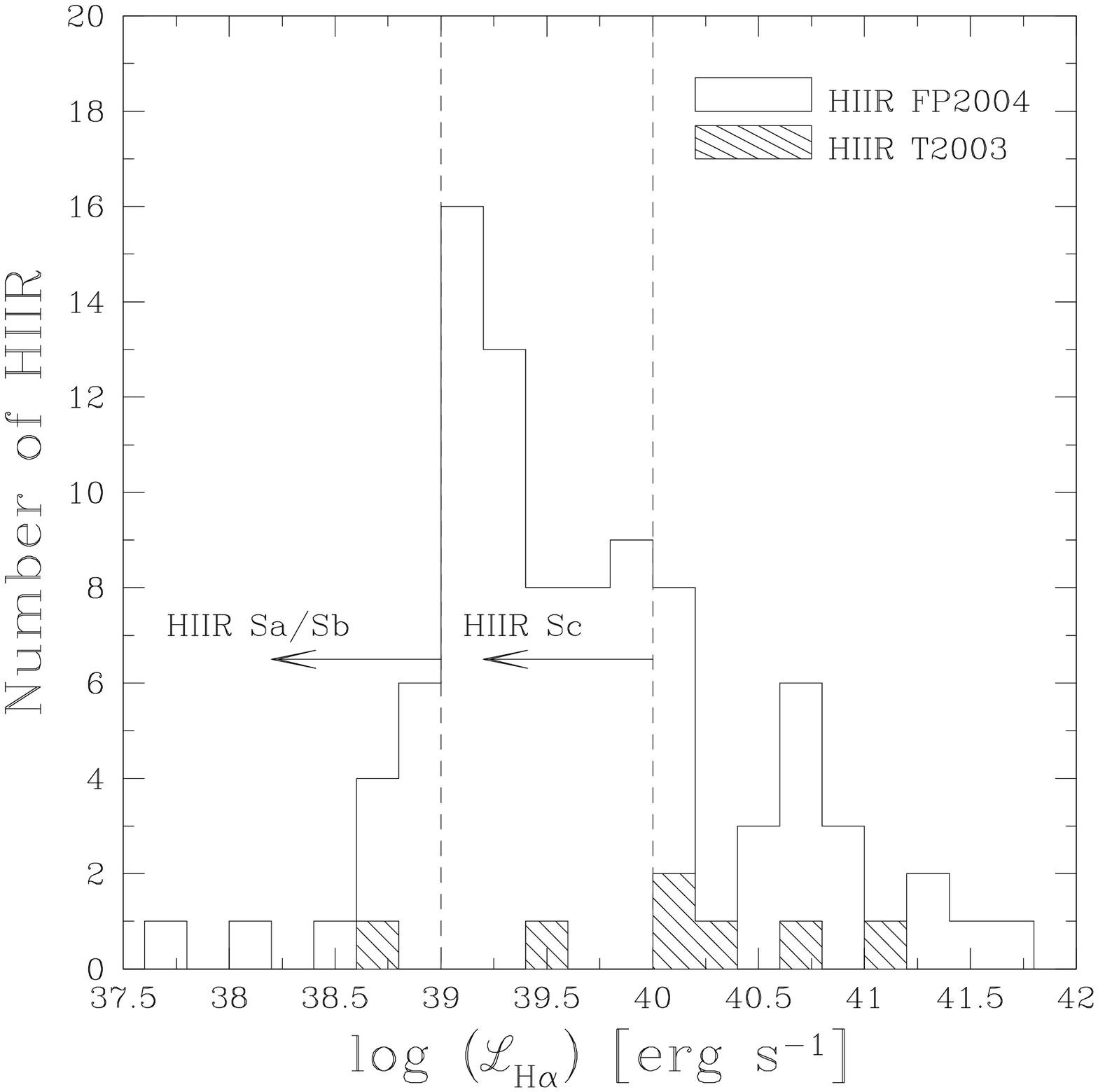}
\caption{Left panel: Distribution of luminosity of the star-forming regions of the sample. TDGs - Tidal Dwarf Galaxies; GHIIC - Giant HII Complex; SSC - Super Stellar Cluster. Right panel: Distribution of H$\alpha$ luminosity of the star-forming regions of our sample (broken line) and the sample of Temporin (2003) (continuous line and shadow area). The horizontal arrows indicate the limits in H$\alpha$ luminosity of HII regions of normal galaxies with same morphological type.}
\label{TDG} 
\end{center}
\end{figure*}

The Blue luminosity histogram Fig.~\ref{TDG} (left panel) shows that 61.5\% of the HII region have $M_B \leq -15\ mag$ (shaded region), 5 of them have magnitude $M_B \leq -19\ mag$, i.e. between 6 to 12 times brighter than the tidal dwarf galaxies candidates of the sample of Weilbacher et al. (2000). The magnitude limit for this sample is $M_B = -17.1\ mag$. Moreover, Mirabel et al. (1992) and Duc \& Mirabel (1994) have confirmed the presence of tidal dwarf galaxies in the extreme of the tidal tale of Arp 105 brighter than $M_B> -19\ mag$. This value is in the order of the most luminous TDGs candidates of our sample. It is important to point out that the TDGs are more luminous (in M$_B$) if compared with star-forming regions in normal galaxies. For example, Weilbacher et al. (2000) find that the TDGs candidates are, on average, $4\ mag$ more luminous than the HII region reported by Bresolin \& Kennicutt (1997). Concerning the $H\alpha$ luminosity of most regions of our sample, they are much more luminous than the most luminous regions observed in spiral galaxies. Star-forming regions with luminosity brighter than $\cal L$$(H\alpha) = 10^{39}\ erg\ s^{-1}$ ($M_V \sim -12.5\ mag$) are not observed in normal Sa and Sb and with $\cal L$$(H\alpha) = 10^{40}\ erg\ s^{-1}$ ($M_V \sim -14\ mag$) in Sc galaxies.  (Bresolin \& Kennicutt 1997).\\

The H$\alpha$ luminosity histogram of our HII region sample (white histogram) and the TDGs sample of Temporin\footnote{Study detailed of the group ultracompact CG J1720-67.8 which contains candidates to TDGs} (2003) (shade histogram) is shown in Fig.~\ref{TDG} (right panel). The histogram was plotted taken bins of log($\cal L(H\alpha)$) = 0.2 ([erg\ s$^{-1}$]). The dotted line indicate the limits in H$\alpha$ luminosity of HII region in normal Sa-Sc galaxies. Twenty six HII region of our sample have H$\alpha$ luminosity brighter than $10^{40}\ erg\ s^{-1}$\\

\begin{figure*} 
\begin{center} 
\includegraphics[width=8cm,height=8cm]{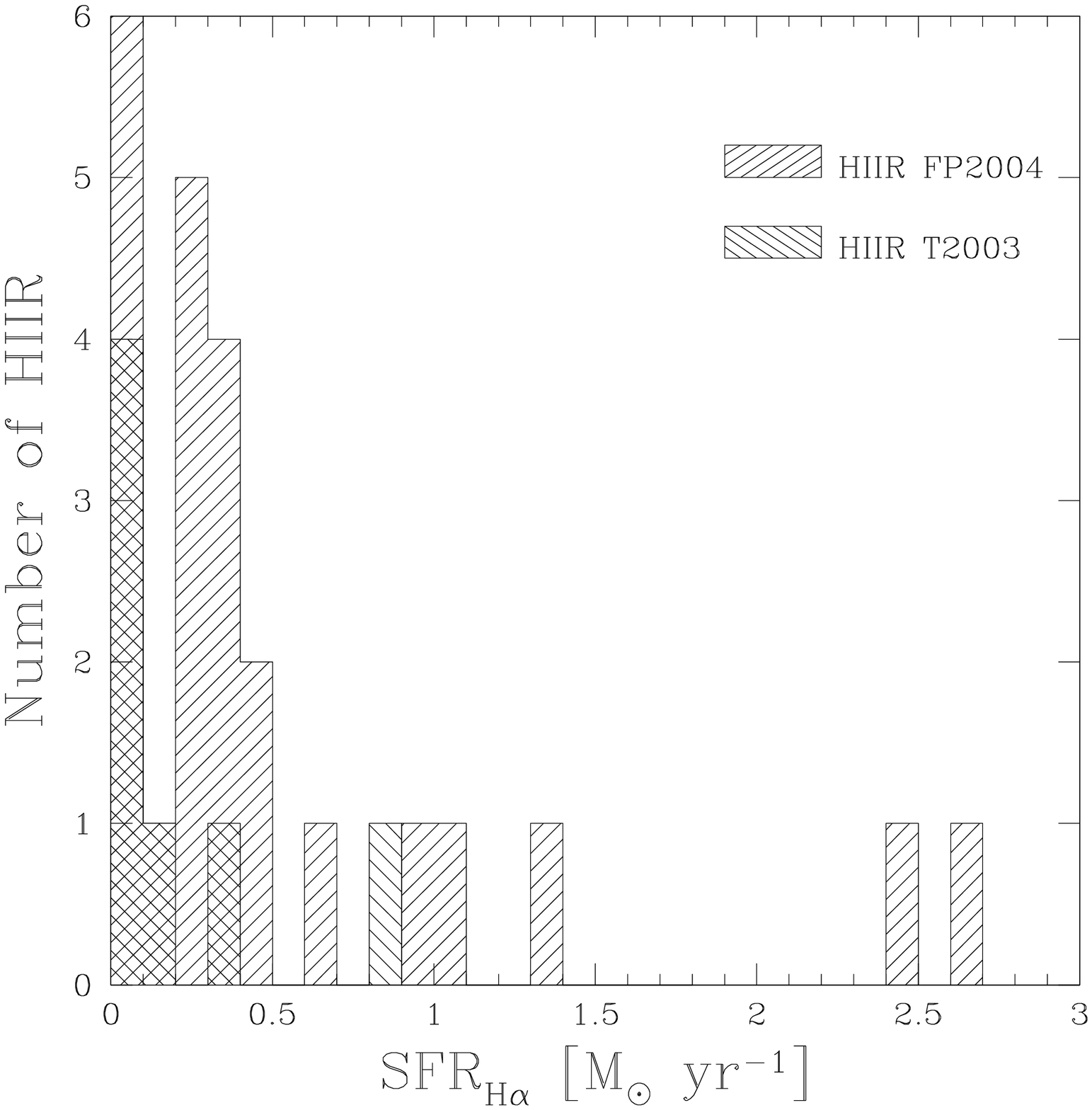}
\includegraphics[width=8cm,height=8cm]{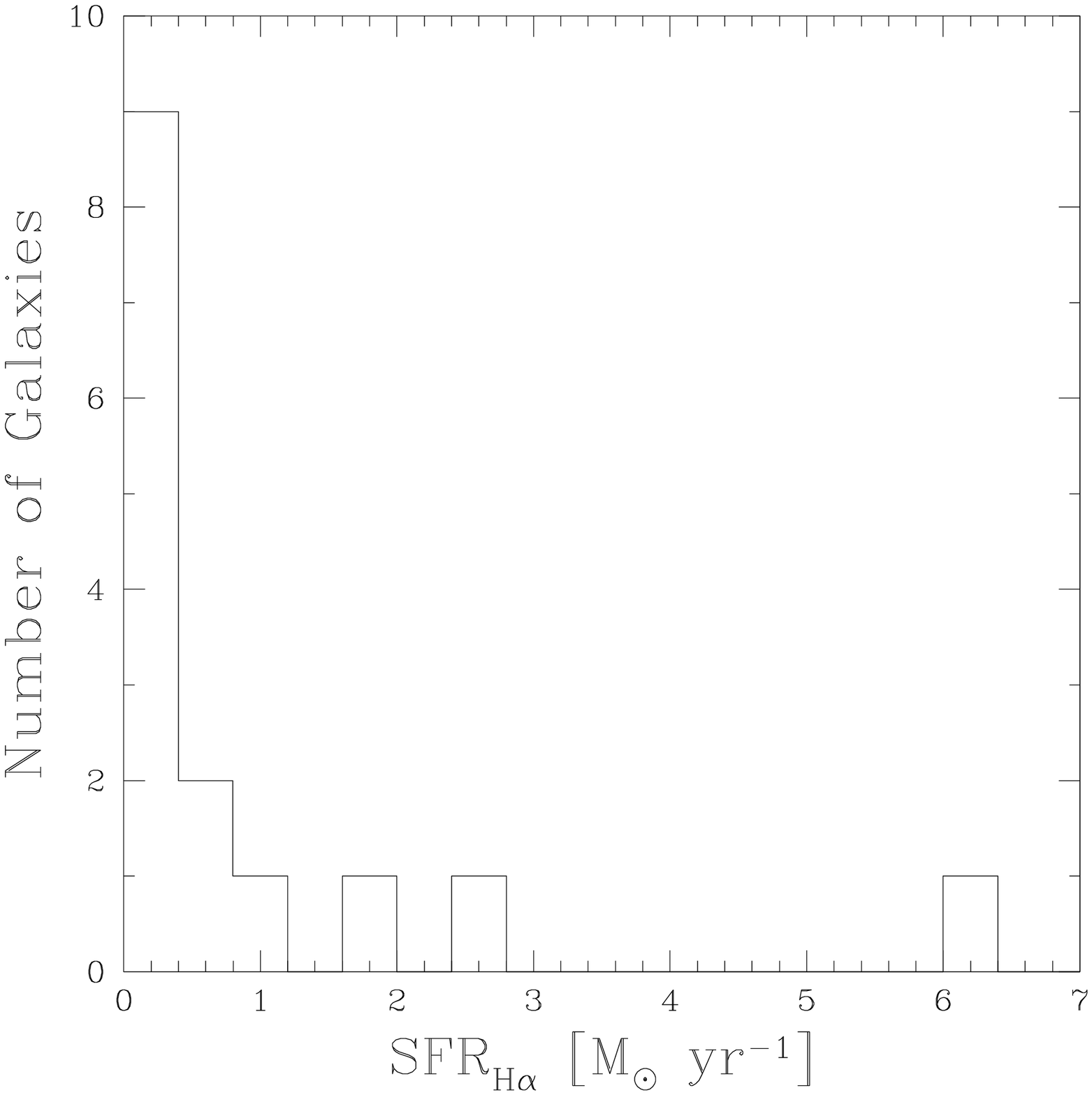}
\caption{Left panel: Distribution of the star formation reckoned of the brightness $H\alpha$ for the HII region of the sample (left shade) and for the HII region of the sample of Temporin (2003) (right shade). The histogram was make taken bins of $SFR= 0.1\ M_\odot\ yr^ {- 1}$. The first bin ($SFR < 0.1\ M_\odot\ yr^ {- 1}$) contain 75 HII regions, by reasons of convenient we do not present the first bin in all its value. Right panel: Distribution of the ``global" star formation reckoned of the brightness $H\alpha$ of the galaxies of the sample. The bins was taken $SFR= 0.4\ M_\odot\ yr^ {- 1}$.} 
\label{SFR TDG} 
\end{center} 
\end{figure*}

Fig.~\ref{SFR TDG} (left panel) presents the distribution of the SFR, estimated from the equation~\ref{SFR(Ha)}, (left shade) and for Temporin´s HII sample (2003) (right shade). The histogram was plotted taken bins of $SFR = 0.1\ M_\odot\ yr^{- 1}$. The first bin ($SFR < 0.1\ M_\odot\ yr^{- 1}$) contain 75 regions. A total of 18 HII regions have $SFR > 0.1\ M_\odot\ yr^{-1}$. Moreover, we can see that 5 HII regions have $SFR> 0.9\ M_\odot\ yr^{- 1}$, this $SRF$ is higher than that of Temporin's sample (2003). In Fig.~\ref{SFR TDG} (right panel) we present the distribution of the ``global" $SFR$ for the galaxies of the sample. This rate was obtained adding the SFR of all the HII regions in each galaxy. The bin is $SFR= 0.4\ M_\odot\ yr^ {- 1} $. This ``global" $SFR$ is the lower limit of SFR in the galaxy since we have inconsiderate the diffuse H$\alpha$ emission. Six galaxies of the sample presents SFR larger than 0.4 $M_\odot\ yr^ {- 1}$.\\

\begin{table}[h!]
\renewcommand{\tabcolsep}{0.5mm}
\caption{Tidal dwarf galaxies candidates: Region, Arp galaxy, blue absolute magnitude, integrated colours, luminosity $H\alpha$, 
number of ionizing photons, age and $SFR$.}
\tiny
\label{TDG-cand}
\begin{center}
\begin{tabular}{ccccccccc} \hline \hline
{Reg.} & {Galaxy} & {$M_B$} & {B-V} & {V-I} & {$\cal L\rm(H\alpha)$} & {$Q(H)$} & {Age} &{SFR($H\alpha$)} \\
&&$[mag]$&&&$[erg\ s^{-1}]$&[phot s$^{-1}$]&[Myrs]&$[M_\odot\ yr^{-1}]$\\
\hline
1  & AM1256E  & -19.80 & 1.61 & 1.96 &---      &---      &13.1 &---\\
N  & AM1401N  & -19.80 & 0.55 & 0.74 &6.13$\times10^{40}$&1.81$\times10^{53}$ &10.2 &3.9$\times10^{-1}$\\
3  & AM1401N  & -19.78 & 0.47 & 0.62 &5.39$\times10^{40}$&1.59$\times10^{53}$ &10.1 &3.4$\times10^{-1}$\\
N  & AM2229W  & -19.49 & 0.66 & 1.21 &1.60$\times10^{41}$&4.71$\times10^{53}$ &7.1  &1.0\\
N  & AM2058S  & -19.24 & 0.27 & 0.47 &5.76$\times10^{40}$&1.79$\times10^{53}$ &6.6  &3.6$\times10^{-1}$\\
1  & AM2030SW & -18.77 & 0.53 & 1.05 &6.72$\times10^{40}$&2.15$\times10^{54}$ &6.5  &4.2$\times10^{-1}$\\
N  & AM2105SE & -18.70 & 0.92 & 1.65 &4.48$\times10^{37}$&1.16$\times10^{53}$ &8.4  &2.8$\times10^{-4}$\\
N  & AM2229E  & -18.69 & 0.56 & 1.12 &1.51$\times10^{41}$&4.50$\times10^{53}$ &6.4  &9.5$\times10^{-1}$\\
2  & AM2030NE & -18.61 & 0.60 & 1.15 &1.12$\times10^{39}$&1.56$\times10^{54}$ &6    &7.1$\times10^{-3}$\\
3  & AM2306N  & -18.55 & 0.42 & 0.59 &4.13$\times10^{41}$&1.21$\times10^{54}$ &6.6  &2.6\\
1  & AM2306N  & -18.50 & 0.45 & 0.83 &3.80$\times10^{41}$&1.13$\times10^{54}$ &6.8  &2.4\\
6  & AM2229W  & -18.44 & 0.61 & 1.16 &4.28$\times10^{40}$&1.27$\times10^{53}$ &7    &2.7$\times10^{-1}$\\
2  & AM1401N  & -18.36 & 0.33 & 0.70 &9.52$\times10^{40}$&2.82$\times10^{53}$ &6.2  &6.0$\times10^{-1}$\\
N  & AM2058N  & -18.18 & 0.99 & 1.14 &6.40$\times10^{39}$&1.96$\times10^{52}$ &7.5  &4.0$\times10^{-2}$\\
2  & AM2306N  & -18.16 & 0.50 & 1.00 &2.16$\times10^{41}$&6.35$\times10^{53}$ &6.7  &1.4\\
4  & AM2058N  & -18.14 & 0.33 & 0.65 &1.80$\times10^{39}$&5.53$\times10^{52}$ &6.2  &1.1$\times10^{-2}$\\
N  & AM2238E  & -18.07 & 1.24 & 1.90 &---      &---      &12.8 &---\\
3  & AM2030NE & -17.98 & 0.61 & 1.17 &8.80$\times10^{39}$&3.41$\times10^{53}$ &6.4  &5.6$\times10^{-2}$\\
5  & AM2229W  & -17.96 & 0.72 & 1.24 &4.76$\times10^{40}$&1.39$\times10^{53}$ &6.9  &3$\times10^{-1}$\\
N  & AM2105NW & -17.90 & 0.63 & 1.57 &3.48$\times10^{38}$&1.16$\times10^{52}$ &6.5  &2.2$\times10^{-3}$\\
N  & AM2330NE & -17.90 & 0.80 & 1.48 &---      &---      &15.3 &---\\
1  & AM2229W  & -17.82 & 0.76 & 1.28 &4.42$\times10^{40}$&1.30$\times10^{53}$ &6.9  &2.8$\times10^{-1}$\\
5  & AM2058N  & -17.80 & 0.41 & 0.75 &1.12$\times10^{40}$&3.49$\times10^{52}$ &6.3  &7.1$\times10^{-2}$\\

1  & AM2030NE & -17.69 & 0.84 & 1.30 &1.28$\times10^{38}$&3.49$\times10^{53}$ &6.4  &8.1$\times10^{-4}$\\
N  & AM2306E  & -17.63 & 0.62 & 1.28 &3.18$\times10^{40}$&8.98$\times10^{52}$ &7.4  &2$\times10^{-1}$\\
10 & AM2058N  & -17.56 & 0.58 & 0.87 &6.64$\times10^{39}$&2.01$\times10^{52}$ &3.6  &4.2$\times10^{-2}$\\
6  & AM1256W  & -17.54 & 0.81 & 1.38 &1.44$\times10^{40}$&4.20$\times10^{52}$ &6.4  &9.1$\times10^{-2}$\\
9  & AM2058N  & -17.41 & 0.50 & 0.77 &6.32$\times10^{39}$&1.92$\times10^{52}$ &3.7  &4.0$\times10^{-2}$\\
3  & AM2238W  & -17.40 & 0.78 & 1.70 &---      &---      &6.1  & ---\\
4  & AM2229W  & -17.38 & 0.70 & 1.17 &3.48$\times10^{40}$&1.03$\times10^{53}$ &6.8  &2.2$\times10^{-1}$\\
5  & AM1256W  & -17.36 & 0.77 & 1.43 &1.43$\times10^{40}$&4.20$\times10^{52}$ &6.1  & 9.0$\times10^{-2}$\\
8  & AM2058N  & -17.34 & 0.46 & 0.75 &5.68$\times10^{39}$&1.71$\times10^{52}$ &6.5  & 3.6$\times10^{-2}$ \\
2  & AM2229W  & -17.31 & 0.55 & 1.06 &7.87$\times10^{40}$&2.31$\times10^{53}$ &6.5  & 5.0$\times10^{-1}$ \\
6  & AM2238W  & -17.30 & 1.03 & 1.97 &---      &---      &6.2  & ---     \\
N  & AM1401S  & -17.30 & 0.71 & 0.92 &4.80$\times10^{39}$&1.44$\times10^{52}$ &7.5  & 3.0$\times10^{-2}$ \\
6  & AM2306E  & -17.25 & 0.49 & 0.93 &3.38$\times10^{40}$&9.62$\times10^{52}$ &6.6  & 2.1$\times10^{-1}$ \\
4  & AM1401N  & -17.12 & 0.47 & 0.72 &1.51$\times10^{40}$&4.47$\times10^{52}$ &6.6  & 9.5$\times10^{-2}$ \\
7  & AM2058N  & -17.12 & 0.44 & 0.70 &6.32$\times10^{39}$&1.92$\times10^{52}$ &6.3  & 4.0$\times10^{-2}$ \\
11 & AM2058N  & -17.10 & 0.58 & 0.86 &3.32$\times10^{39}$&1.01$\times10^{52}$ &4.0  & 2.1$\times10^{-2}$ \\
N  & AM2322SE & -16.99 & 1.18 & 1.22 &1.28$\times10^{40}$ &3.83$\times10^{52}$ &7.3  & 8.1$\times10^{-2}$ \\
1  & AM2238W  & -16.90 & 1.03 & 1.91 &---      & ---      &6.4  &---     \\
4  & AM1256W  & -16.74 & 0.76 & 1.31 &5.94$\times10^{39}$ &1.75$\times10^{52}$ &5.9  & 3.8$\times10^{-2}$ \\
1  & AM1401N  & -16.73 & 0.31 & 0.66 &1.77$\times10^{40}$ &5.23$\times10^{52}$ &6.2  & 1.1$\times10^{-1}$ \\
x  & AM2238W  & -16.70 & 0.96 & 1.87 &---      & ---      &6.3  &---     \\
2  & AM1256E  & -16.70 & 1.11 & 1.73 &---      & ---      &7.0  &---     \\
9  & AM2238W  & -16.60 & 0.85 & 1.85 &---      & ---      &6.4  &---     \\
1  & AM1448SW & -16.52 & 0.76 & 1.30 &---      & ---      &7.1  &---     \\
8  & AM2238W  & -16.30 & 0.99 & 1.80 &---      & ---      &6.3  &---     \\
7  & AM2306E  & -16.26 & 0.48 & 0.92 &1.09$\times10^{40}$ &3.11$\times10^{52}$ &6.6  & 6.9$\times10^{-2}$ \\
1  & AM2306E  & -16.24 & 0.48 & 0.85 &1.41$\times10^{40}$ &4.01$\times10^{52}$ &6.5  & 8.9$\times10^{-2}$ \\
3  & AM2229W  & -16.18 & 0.71 & 1.18 &3.60$\times10^{39}$ &1.05$\times10^{52}$ &7.2  & 2.3$\times10^{-2}$\\
1  & AM1256W  & -16.12 & 0.69 & 1.26 &3.28$\times10^{39}$ &9.62$\times10^{51}$ &6.3  & 2.1$\times10^{-2}$ \\
3  & AM1256W  & -16.10 & 0.76 & 1.28 &4.37$\times10^{39}$ &1.30$\times10^{52}$ &6.0  & 2.8$\times10^{-2}$ \\
2  & AM1256W  & -15.83 & 0.81 & 1.28 &4.52$\times10^{39}$ &1.33$\times10^{52}$ &6.3  & 2.9$\times10^{-2}$ \\
9  & AM2306E  & -15.76 & 0.48 & 0.91 &7.37$\times10^{39}$ &2.10$\times10^{52}$ &6.6  & 4.7$\times10^{-2}$ \\
N  & AM1448NE & -15.73 & 1.31 & 1.63 &---      & ---      &13.7 &---     \\
5  & AM2238W  & -15.70 & 0.78 & 1.76 &---      & ---      &6.2  &---     \\
7  & AM1256W  & -15.69 & 0.62 & 1.05 &2.44$\times10^{39}$ &7.13$\times10^{51}$ &4.2  & 1.5$\times10^{-2}$ \\
10 & AM2306E  & -15.63 & 0.48 & 0.88 &4.18$\times10^{39}$ &1.18$\times10^{52}$ &6.7  & 2.6$\times10^{-2}$ \\
2  & AM2030SW & -15.60 & 0.61 & 1.02 &6.72$\times10^{39}$ &1.36$\times10^{53}$ &6.5  & 4.2$\times10^{-2}$ \\
1  & AM2322NW & -15.52 & 0.67 & 0.76 &1.07$\times10^{40}$ &2.71$\times10^{52}$ &7.3  & 6.8$\times10^{-2}$ \\
x  & AM2238W  & -15.50 & 1.04 & 1.80 &---      & ---      &6.5  &---     \\
7  & AM2238W  & -15.50 & 0.89 & 1.78 &---      & ---      &6.2  &---     \\
2  & AM1448SW & -15.45 & 0.85 & 1.42 &---      &  ---      &7.1  &---     \\
8  & AM2306E  & -15.42 & 0.48 & 0.91 &7.64$\times10^{39}$ &2.15$\times10^{52}$ &6.5  & 4.8$\times10^{-2}$ \\
6  & AM2058N  & -15.41 & 0.44 & 0.50 &1.89$\times10^{39}$ &5.80$\times10^{51}$ &5.1  & 1.2$\times10^{-2}$ \\
3  & AM2058N  & -15.40 & 0.31 & 0.62 &1.96$\times10^{39}$ &5.93$\times10^{51}$ &5.0  & 1.2$\times10^{-2}$ \\
2  & AM2058N  & -15.35 & 0.41 & 0.47 &2.50$\times10^{39}$ &7.64$\times10^{51}$ &5.6  & 1.6$\times10^{-2}$ \\
4  & AM2238W  & -15.20 & 0.93 & 1.86 &---      & ---      &6.1  &---     \\
2  & AM2306E  & -15.15 & 0.50 & 0.97 &4.09$\times10^{39}$ &1.16$\times10^{52}$ &6.7  & 2.6$\times10^{-2}$ \\
1  & AM2058N  & -15.13 & 0.32 & 0.67 &1.91$\times10^{39}$ &7.24$\times10^{51}$ &5.0  & 1.2$\times10^{-2}$ \\
7  & AM2238W  & -15.50 & 0.89 & 1.78 &---      & ---      &6.2  &---    \\
2  & AM2238W  & -15.00 & 0.82 & 1.87 &---      & ---      &6.2  &---     \\
\hline \hline
\end{tabular}
\end{center}
\end{table}

\section{Modeling the ionizing cluster}
\label{modeling the Ionizing cluster}

In order to obtain a better insight on the internal properties that govern the star formation process in the HII regions, we have compared the number of ionizing photons estimated from the measured $\cal L\rm (H\alpha)$ luminosity listed in Table~\ref{TDG-cand}, with that of models of an ionizing cluster.\\

Two observational constrain were used for the models:  
\begin{itemize}
\item\ 1) The age of the HII region is older than $3.6 \times 10^6$ yrs, therefore the turn off point of the cluster is at O5 spectral type.
\item\ 2) The number of ionizing photons is $Q(H) > 10^{51}$ photon s$^{-1}$. 
\end{itemize}

We calculated for the cluster models, the total number of UV photons ($Q(H)$), the number of stars in spectral type and the cluster mass adopting for the cluster the Salpeter's IMF with $\alpha = -2.65$ and Kurucz's stellar atmosphere models.\\

The results for three ionizing cluster models are present in Table~\ref{table 2}. The stellar spectral type is listed in column 1. In column 2 we list the number of stars for each spectral type. The total number of ionized photons is listed in column 3 and column 4 gives the total mass lacked up in each spectral type. In order to reproduce $Q(H) = 1.58 \times 10^{51}$ ionizing photons required for the upper limit of the Salpeter's IMF two stars O5 of 60 M$_{\odot}$. $Q(H) = 4.8 \times 10^{52}$ photons s$^{-1}$ would require 60 of such stars, while for $Q(H) = 4.0 \times 10^{53}$ photons s$^{-1}$ 500 O5 stars would be necessary. The total lower limit of the stellar mass for each cluster models are $0.2 \times 10^6$, $2.4  \times 10^6$ and $5.2 \times 10^7$ respectively. These values agree with young massive clusters (10$^4 \leq M/M_\odot \leq 10^7$) found in the outer regions of the galactic disk or in the tidal tails of the NGC\,6872 (Bastian et. al 2005).

\renewcommand{\tabcolsep}{4mm}
\begin{table}[ht]
\caption{Ionizing cluster models: Spectral type, number of stars, number of ionized photons and the total mass correspondent to each spectral type. }
\begin{center}
\begin{tabular} {c|r r r}
\hline
\hline
\multicolumn{4}{c}{Cluster No 1}\\
\hline
\multicolumn{1}{c|}{Spectral}	&\multicolumn{1}{c}{Stars}	&\multicolumn{1}{c}{$Q(H)$}	&\multicolumn{1}{c}{Mass} \\
\multicolumn{1}{c|}{Type}	&\multicolumn{1}{c}{[No]}	&\multicolumn{1}{c}{[phot s$^{-1}$]}	&\multicolumn{1}{c}{[$M_\odot$]}\\
\hline
 O5	& 2        &  $1.50\times 10^{51}$  & $120  $ \\
 O6	& 14       &  $5.37\times 10^{49}$  & $518  $ \\
 O8	& 33       &  $2.97\times 10^{49}$  & $759  $ \\
 B0	& 212      &  $1.60\times 10^{49}$  & $3710 $ \\
 B5	& 687      &  $5.48\times 10^{47}$  & $4053 $ \\
 AO	& 9362     &  $4.10\times 10^{41}$  & $27149$ \\
 F0	& 14605    &         -              & $23368 $ \\
 G0	& 25289    &         -              & $27818 $ \\
 K0	& 51832    &         -              & $46689 $ \\
 M0	& 146397   &         -              & $73198 $ \\ 
\hline                                        
TOTAL	& 233828   &  $1.58\times 10^{51}$  & $207382$ \\
\hline
\multicolumn{4}{c}{}\\
\hline
\hline
\multicolumn{4}{c}{Cluster No 2}\\
\hline
 O5	& 60         &  $4.50\times 10^{52}$  & 3600    \\
 O6	& 437        &  $1.68\times 10^{51}$  & 16169   \\
 O8	& 986        &  $8.87\times 10^{50}$  & 22678   \\
 B0	& 6395       &  $4.81\times 10^{50}$  & 108715  \\
 B5	& 20617      &  $1.64\times 10^{49}$  & 123702  \\
 AO	& 103311     &  $4.52\times 10^{42}$  & 299602  \\
 F0	& 161170     &         -              & 257872  \\
 G0	& 279068     &         -              & 306974  \\
 K0	& 571980     &         -              & 457584  \\   
 M0	& 1615518    &         -              & 807759  \\
\hline	
TOTAL	& 2760542    &  $4.80\times 10^{52}$  & 2404655 \\
\hline

\multicolumn{4}{c}{}\\
\hline
\hline
\multicolumn{4}{c}{Cluster No 3}\\
\hline
 O5	& 500        &  $3.75\times 10^{53}$  & 30000    \\
 O6	& 3659       &  $1.40\times 10^{52}$  & 135272   \\
 O8	& 8235       &  $7.41\times 10^{51}$  & 189405   \\
 B0	& 53305      &  $4.01\times 10^{51}$  & 906185   \\
 B5	& 171809     &  $1.37\times 10^{50}$  & 1030854  \\
 AO	& 2330741    &  $1.02\times 10^{44}$  & 6992223  \\
 F0	& 3636075    &         -              & 5817720  \\
 G0	& 6295907    &         -              & 6925498  \\
 K0	& 12904121   &         -              & 11613709 \\   
 M0	& 36446801   &         -              & 18223400 \\
\hline	
TOTAL	& 132806     &  $4.00\times 10^{53}$  & 51864266 \\
\hline
\end{tabular}
\end{center}
\label{table 2}
\end{table}

\section{Conclusions}
\label{discussions and conclusions}

We have presented an analysis of the nature of 117 star formation regions, which were observed in 11 minor mergers of galaxies. We classify these regions using three specific parameters: i) the blue absolute magnitude $M_B$; ii) the $H\alpha$ luminosity $\cal L\rm (H\alpha)$ and iii) the star formation rate $SFR_{H\alpha}$. The properties of these regions were compared with the other TDGs candidates. We have found that 61.5\% of the HII regions have $M_B \leq -15\ mag$, which is the upper limit of Giants HII Complex (GHIIC) observed by Bresolin \& Kennicut (1997) in Sa/Sb type galaxies. Therefore we conclude that most of these are TDGs candidates. We point out that the first 5 regions listed in table 1 are six times brighter than the tidal dwarf galaxies candidates of the Weilbacher et al. (2000) sample, and as bright as that observed in the tidal tale of the Arp 105 system. We have also found that 26 regions have H$\alpha$ luminosities brighter than $10^{40}\ erg\ s^{-1}$, which is the upper limit in luminosity for HII regions in normal Sc galaxies.\\

We found that 22 HII regions have $SFR > 0.1\ M_\odot\ yr^{-1}$, this value is typical of TDGs. Very high SFR is found for 5 HII regions: $SFR > 0.9\ M_\odot\ yr^{-1}$, this $SRF$ is higher than in the sample of Temporin (2003). \\

We have compared the number of ionizing photons estimated from the measured $H\alpha$ luminosity ($\cal L\rm (H\alpha)$), with that given by models of ionizing clusters. Using two observational constrain for the models we calculated the total number of UV photons of the cluster models, the number of the star of each spectral type and the total stellar mass, adopting for the cluster the Salpeter's luminosity function with $\alpha = -2.65$ and the Kurucz's stellar atmosphere models. We found that in order to reproduce the total number of the ionizing photons required for the upper limit of the IMF is between 2 to 500 05 stars of 60 solar masses. The estimated total lower mass limits for the three cluster models are $0.2 \times 10^6$, $2.4 \times 10^6$ and $5.1 \times 10^7$ respectively. These values suggest that TDGs candidates   would be more massive than   young massive clusters found in the outer regions of the galactic disk or the tidal tails of the NGC\,6872.\\

\begin{acknowledgements}
This work has been partially supported by the CONICET and The Brazilian Institution CNPq (PRONEX 66.2088/1997-2). \\
\end{acknowledgements}

\end{document}